\documentclass{article}

\usepackage{amssymb}
\usepackage{amsmath}
\usepackage{graphicx}
\usepackage{epstopdf}

\newcommand{\pfrac}[2]{\frac{\partial #1}{\partial #2}}
\newcommand{\Hilb}{\mathcal{H}}
\newcommand{\Four}{\mathcal{F}}
\newcommand{\inftyint}{\int_{-\infty}^\infty}
\newcommand{\piint}{\int_{-\pi}^\pi}
\DeclareMathOperator{\sign}{sign}
\usepackage{physics}

\usepackage{authblk}

\begin{document}

\title{Mesh density functions based on local bandwidth applied to moving mesh methods}
\author{Elliott~S.~Wise\thanks{Corresponding author: elliott.wise.14@ucl.ac.uk}}
\author{Ben~T.~Cox\thanks{b.cox@ucl.ac.uk}}
\author{Bradley~E.~Treeby\thanks{b.treeby@ucl.ac.uk}}

\affil{Department of Medical Physics and Biomedical Engineering,
	University College London\\
	2--10 Stephenson Way, London, NW1 2HE, United Kingdom}


\maketitle

\begin{abstract}
	Moving mesh methods provide an efficient way of solving partial differential equations for which large, localised variations in the solution necessitate locally dense spatial meshes.
	In one-dimension, meshes are typically specified using the arclength mesh density function.
	This choice is well-justified for piecewise polynomial interpolants, but it is only justified for spectral methods when model solutions include localised steep gradients.
	In this paper, one-dimensional mesh density functions are presented which are based on a spatially localised measure of the bandwidth of the approximated model solution.
	In considering bandwidth, these mesh density functions are well-justified for spectral methods, but are not strictly tied to the error properties of any particular spatial interpolant, and are hence widely applicable.
	The bandwidth mesh density functions are demonstrated by applying periodic spectral and finite-difference moving mesh methods to a number of model problems in acoustics.
	These problems include a heterogeneous advection equation, the viscous Burgers' equation, and the Korteweg-de Vries equation.
	Simulation results demonstrate solution convergence rates that are up to an order of magnitude faster using the bandwidth mesh density functions than uniform meshes, and around three times faster than those using the arclength mesh density function.
\end{abstract}

\section{Introduction}

Many scientific and engineering problems require solutions to partial differential equations (PDEs).
When smooth, these solutions can be efficiently computed using spectral methods.
However, often solutions are not equally smooth everywhere.
In particular, they might exhibit features which are tightly localised in space.
These include shock fronts, narrow pulses, and sharp corners.
Such features require dense computational meshes to accurately resolve.
Because spectral methods typically use standardised meshes, the global mesh density is determined by the sampling requirements of these localised features.
This results in much of the spatial domain being oversampled, increasing computational expense for no accuracy gain.
As an example of when this can become a critical issue, three-dimensional, full-wave simulations of nonlinear medical ultrasound fields may require many tens of gigabytes of memory to store acoustic field variables at each time-step due to large, densely sampled simulation domains \cite{Jaros2016}.
These sampling requirements arise when acoustic nonlinearity causes very high frequencies to form, often within small regions where the acoustic pressure is particularly high.

Adaptive moving mesh methods can reduce the tradeoff between accuracy and computational expense by providing more optimal sampling.
They place mesh nodes according to a monitor function (sometimes called a mesh density function in one dimension) that is computed from (and locally dependent on) the calculated solution itself.
Moving mesh methods have traditionally been implemented using finite-difference and finite-element methods, but spectral implementations offer the opportunity to improve computational efficiency further.
Some examples of spectral moving mesh methods include Fourier \cite{Feng2006,Feng2009}, Galerkin \cite{Shen2009}, and Chebyshev \cite{Tapia2009} types.
These all used the arclength monitor function, which clusters mesh nodes according to the gradient of the model solution.
For these problems, this choice is justified by physical considerations: the model solutions in all cases feature localised steep gradients.
However, it is not clear why the arclength monitor function might produce a mesh that is optimal.

One justification for the arclength monitor function is given in \cite[\S2.4]{Huang2011}.
Here, it is shown that derivative--based monitor functions can be derived from interpolation error bounds for piecewise polynomial interpolants.
The arclength monitor function, while not strictly optimal, can be seen to be very similar to the optimal monitor functions they derive.
But these optimal monitor functions do not naturally extend to spectral interpolants.
One approach to deriving optimal meshes for spectral methods has been to directly consider smoothness properties of the approximated solution itself.
A notable one-dimensional example is found in the work of Tee et al.\ \cite{Tee2006a,Tee2006b,Hale2009a,Hale2009b}.
Their approach is designed for solutions whose analytic continuations contain singularities.
It works by first approximating the analytic continuation, after which a mesh mapping is computed that is parametrised by the locations of the approximated singularities.
These mesh mappings seek to ensure that a spectral interpolant through the composition of the approximated solution and inverse mesh mapping converges on the true solution faster than a spectral interpolant through the solution alone.
In \cite{Wise2015}, a mesh density function is presented which is based on Tee et al.'s approach.
This work demonstrated that a singularity--based mesh density function was significantly more effective than the arclength mesh density function in reducing the tradeoff between accuracy and computational expense.
However, an obvious limitation of this approach is that it requires the model solution's analytic continuation to include singularities (or at least near--singular behaviour).

Recently, Subich \cite{Subich2015} demonstrated a more general one-dimensional spectral moving mesh method, which uses a mesh density function that is given by the envelope of the high-frequency components of the model solution.
However, no direct justification is provided for why a solution envelope should correspond to a beneficial mesh density.
Moreover, the threshold beyond which frequencies are considered to be high is chosen based on model-specific and interpolant-specific considerations.
Nonetheless, using the local frequency content of a solution to form a mesh density function is an intuitive and general approach that has remained largely unexplored.

This paper presents a family of mesh density functions that are based on the local bandwidth of the model solution.
This approach is justified by a nonuniform analogue of the Nyquist sampling theorem, and so does not depend on either problem--specific or interpolant--specific considerations.
This makes them applicable to both spectral and non-spectral methods, as well as a wide range of problem types.
The mesh density functions are presented in this paper through application to a variety of acoustics problems.
These numerical experiments are primarily conducted using a periodic spectral moving mesh method, with some additional validation for periodic finite-difference moving mesh methods.

\section{Bandwidth mesh density functions}

In this section, mesh density functions are derived from the local spatial frequency content of the solution to a model PDE.
For notational purposes, it is useful to consider such a solution as a signal $u$ that is a function of a spatial coordinate $x$ with corresponding wavenumbers $k$.
Sampling criteria for such signals are typically based on spatial frequency considerations.
The most famous is the Nyquist--Shannon sampling theorem for band-limited signals, which states that perfect reconstruction requires a sampling rate of at least twice the signal's bandwidth, which is measured from DC to the maximum frequency present.\footnote{A similar criterion applies to signals whose power is zero below some lower bound on the absolute frequency (non-baseband signals), but this requires explicit knowledge of the lower bound, and so is less general.}
The Nyquist--Shannon sampling criterion is typically used with uniform sampling, but this is not strictly required.
Band-limited signals with a spatially varying frequency content can be perfectly reconstructed using non-uniform samples taken at a rate equal to twice the local bandwidth \cite{Clark1985}.
Of course, not all signals possess a hard cut-off in their frequency content.
Hence, it is useful to consider statistical measures of bandwidth.

Now let the signal $u$ be $2\pi$-periodic and normalised such that
\[
	\piint|u|^2dx = \inftyint|U|^2dk=1,
\]
where $U$ is the Fourier transform of $u$.
The local amplitude $A$, phase $\varphi$, and spatial frequency $\varphi_x$ (subscripts denote differentiation) of this signal are inherently coupled when the signal is directly analysed.
The analytic signal provides a way of decoupling them.
It does this by attaching an imaginary counterpart which is in quadrature with the signal.
This is typically done via the Hilbert transform $\Hilb$, yielding
\begin{equation}
	\label{eq:analytic_signal}
	v = u+i\Hilb u = Ae^{i\varphi}.
\end{equation}
To define a local, statistical measure of bandwidth, the analytic signal is considered to have a joint position--wavenumber power density $P(k,x)$ which satisfies the marginals
\[
	P(x) = \inftyint P(k,x)dk = |v|^2,\qquad
	P(k) = \piint P(k,x)dx = |V|^2,
\]
where $V$ is the Fourier transform of $v$.
A useful family of statistical bandwidth measures are the even-order spectral moments
\begin{equation}
	\label{eq:power-moment}
	\langle k^{2m}\rangle = \inftyint k^{2m}P(k)dk.
\end{equation}
The order of the spectral moments can be chosen based on the desired weighting that is to be given to the power density's tail.
Since the $m=1$ moment corresponds to the variance (assuming symmetry in $P(k)$ about $k=0$), and the square root of the variance is commonly used as a measure of bandwidth, the second spectral moment is used for the derivation that follows.
Similar derivations are easily made for other choices of $m$.
From the global spectral moment, a local equivalent may be derived using the conditional power density
\[
	P(k|x) = \frac{P(k,x)}{P(x)}
\]
as
\begin{align}
	\label{eq:power-freq2space}
	\langle k^2\rangle 
	&= \inftyint k^2P(k)dk \nonumber\\
	&= \inftyint k^2\left(\piint P(k,x)dx\right)dk \nonumber\\
	&= \inftyint k^2\left(\piint P(k|x)P(x)dx\right)dk \nonumber\\
	&= \piint\left(\inftyint k^2 P(k|x)dk\right)P(x)dx \nonumber\\
	&= \piint\langle k^2\rangle|_xP(x)dx
\end{align}
Here, the local second spectral moment has been defined by \cite{Davidson2000}
\begin{equation}
	\label{eq:local_moment_jointdensity}
	\langle k^2\rangle|_x = \inftyint k^2P(k|x)dk.
\end{equation}
This choice of definition is intuitive when $P(k|x)$ is considered as a local frequency distribution, and when \eqref{eq:power-moment} is compared with \eqref{eq:local_moment_jointdensity}.
While \eqref{eq:local_moment_jointdensity} could be used to compute the local bandwidth of a signal, it is convenient to take an alternative approach that avoids explicitly computing a joint power density.

An alternative way of computing local spectral statistics is to consider the operator
\[
	K = \left\{\begin{array}{c l}
		\frac{1}{i}\frac{d}{dx} & \text{in the position representation} \\
		k                       & \text{in the wavenumber representation.}
	\end{array}\right.
\]
For the wavenumber representation, this operator can be used to compute the global second spectral moment as
\begin{equation}
	\label{eq:globalmoment_wavenumber}
	\langle k^2\rangle=\expval{K^2}{V} = \inftyint k^2|V|^2dk,
\end{equation}
as expected.
Similarly, for the position representation the global second spectral moment is given by \cite{Davidson2000} 
\begin{equation}
	\label{eq:globalmoment}
	\langle k^2\rangle = \expval{K^2}{v}
	= \piint\left|\frac{v_x}{v}\right|^2|v|^2dx.
\end{equation}
Making a comparison with \eqref{eq:power-freq2space}, the left term of the integrand in \eqref{eq:globalmoment} is considered to be the local second spectral moment \cite{Cohen1993,Davidson2000}:
\begin{equation}
	\label{eq:conditionalmoment}
	\langle k^2\rangle|_x = \left|\frac{v_x}{v}\right|^2. 
\end{equation}
In \cite{Cohen1996}, this interpretation of similar operators is justified by showing that it leads to established results for a number of quantum mechanical statistics.
Taking the square root of \eqref{eq:conditionalmoment} then gives a definition of the local bandwidth, which is used to define the first of two proposed mesh density functions: the \emph{ordinary bandwidth mesh density function}
\begin{equation}
	\label{eq:fourmd}
	\rho = \left|\frac{v_x}{v}\right|.
\end{equation}
A complication arises when this mesh density function is computed from an analytic signal $v$ whose amplitude drops to zero, since computing the local bandwidth becomes ill-posed in these regions.
One approach to regularisation is to include an amplitude-weighting.
This leads to the \emph{amplitude-weighted bandwidth mesh density function}, defined as
\begin{equation}
	\label{eq:fourmdaw}
	\rho = |v_x|.
\end{equation}
Both of the bandwidth mesh density functions require an analytic signal to be computed from the solution to the model PDE at a given point in time.
To this end, an algorithm is provided in Appendix~A for computing a Hilbert transform from nonuniform samples of a function.

\section{Numerical experiments}

\subsection{Numerical methods}

To examine the bandwidth mesh density functions, two moving mesh methods are used.
These are outlined below, but it is important to note that the bandwidth mesh density functions are agnostic to the algorithmic choices that have been made.
The two moving mesh methods are differentiated by the numerical method they use for computing spatial gradients in the model/mesh PDEs.
Both assume periodic model solutions, but one uses spectral interpolants and the other uses centered finite-differences.
For the spectral interpolants, gradients are computed using either a collocating Fourier interpolant (when taken with respect to the computational coordinate) or a rational trigonometric interpolant \cite{Baltensperger2002,Tee2006b} (when taken with respect to the physical coordinate).
For the finite-difference method, gradients are computed using centered finite-differences of various accuracy-orders taken with respect to the computational coordinate.
Physical gradients are then computed using the chain rule
\[
	\pfrac{}{x}=\pfrac{}{s}\pfrac{s}{x},
\]
where $x$ and $s$ are the physical and computational coordinates, respectively.
The notation $x=x(t,s)$ is used to refer to $x$ as both the physical coordinate itself, and as the time-varying transformation between the physical and computational coordinates.

Mesh node relocation is controlled using the moving mesh PDE (MMPDE) \cite{Huang1994}
\begin{equation}
\label{eq:mmpde5}
x_t = \tau^{-1}(\rho x_s)_s.
\end{equation}
The function $\rho(x)$ is the solution-dependent mesh density specification, which the MMPDE satisfies as $t\to\infty$.
The rate at which this happens is controlled by the scalar parameter $\tau$, which must match the time-scale over which solution features evolve.
Note that to compute the mesh gradient $x_s$, a slight adjustment must be made to the approaches described in the preceding paragraph.
Gradients are computed via the formula $x_s=(x-s)_s+1$ (assuming $x$ and $s$ share a domain size).
This is done because $x-s$ is periodic for the work presented in this paper, while $x$ is not.
Strictly speaking, this approach may not guarantee monotonicity in the implied continuous mesh mapping, but in practice a well-sampled mesh density function ensures that this is the case.

A mesh density function is usually spatially smoothed before being applied to a MMPDE.
This is done for a number of reasons.
First, it ensures that the mesh density function is well-sampled by the mesh, and hence that the MMPDE can be efficiently solved.
Second, it ensures smooth mesh transformations, which in turn produce fast convergence rates when the model is solved.
Third, it lessens the inherent increase in stiffness that comes with mesh adaptivity, improving the efficiency of time-stepping algorithms \cite{Huang1997}.
Let $\tilde{\rho}$ and $\rho$ be the smoothed and unsmoothed mesh density functions respectively.
These are related by the equation
\begin{equation}
\label{eq:smoothing}
\tilde{\rho}-\beta^{-2}\tilde{\rho}_{ss} = \rho,
\end{equation}
where the parameter $\beta$ controls the degree of smoothing \cite{Huang1997,Huang2011}.
With homogeneous Neumann boundary conditions, \eqref{eq:smoothing} constrains the relative rate of change in the smoothed mesh density by $|\tilde{\rho}_s|/\tilde{\rho}\leq\beta$, and similar behaviour is observed for periodic boundary conditions.
Equation~\eqref{eq:smoothing} is solved using a Fourier interpolant via
\begin{equation}
\label{eq:smoothingdiscrete}
\tilde{\rho} = \Four^{-1}\left\{\frac{\Four\{\rho\}}{1+\beta^{-2}k^2}\right\},
\end{equation}
where $k$ are wavenumbers corresponding to $s$.
The smoothing parameter can be chosen to be discretisation-dependent, so that it has a similar effect to nearest-neighbour smoothing.
This choice aims to ensure that the mesh density function is well-sampled, since, for example, a well-sampled peak will be smoothed (and reduced) less than a poorly sampled one.
Unless otherwise stated, here $\beta=(\Delta s\sqrt{2})^{-1}$ \cite{Huang1997}, where $\Delta s$ is the discretisation size in the computational coordinate.
A numerical experiment is described in Appendix B, which analyses this choice for one of the model problems presented below.

The model and mesh PDEs are coupled together using a quasi-Lagrange approach.
This replaces the usual time-derivatives in the model PDE with an advective derivative
\[
\frac{d}{dt}=\pfrac{}{t}+\pfrac{}{x}\pfrac{x}{t}.
\]
It is notable that this yields an implicit system of PDEs, and that the coupled model/mesh system may be stiffer than either PDE individually.
Time-stepping is performed using the method of lines, with Matlab's \verb|ode15i| function \cite{Shampine2002} used to integrate the system of ordinary differential equations that results from the spatial discretisation previously discussed.
This algorithm solves ODEs of the form $f(t,y,y_t)=0$ using adaptive-order backward differentiation formulae, with an adaptive timestep size.
Here, $y$ is a vector of solution values and mesh node positions, and $y_t=dy/dt$.

\subsection{Example problems}

The bandwidth mesh density functions are demonstrated through application to four problems and three acoustic models.
These each exhibit different feature types.
The first problem is based on a heterogeneous advection equation, and exhibits the formation and propagation of a sharp crest.
The second problem is based on the viscous Burgers' equation, and exhibits the formation of a stationary shock front.
The third problem is also based on the viscous Burgers' equation, and exhibits the formation, propagation, and merging of multiple shock fronts.
The fourth problem is based on the Korteweg-de Vries equation, and exhibits the formation of multiple solitons, and their subsequent interactions.
All use dimensionless units and a periodic domain $x\in[-\pi,\pi)$.
The time-stepping algorithm is provided with relative and absolute error tolerances of $10^{-9}$ and $10^{-10}$, to ensure that errors in the spatial numerical method dominate.
The mesh speed parameter is $\tau=10^{-2}$.
For the illustrations presented in Figs.~\ref{fig:advection}--\ref{fig:complicated-KdV} below, the amplitude-weighted bandwidth mesh density function was used in conjunction with the spectral moving mesh method described in \S3.1.

\subsubsection{A heterogeneous advection equation}

\begin{figure}
	\includegraphics[width=\textwidth]{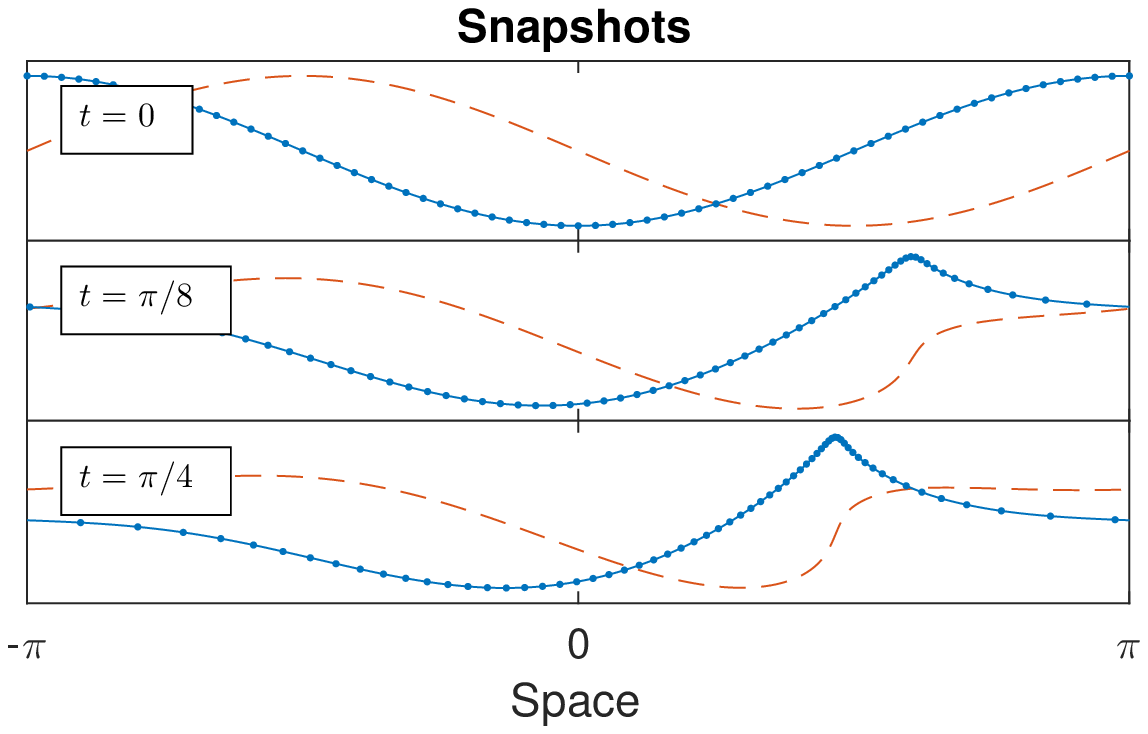} 
	\includegraphics[width=\textwidth]{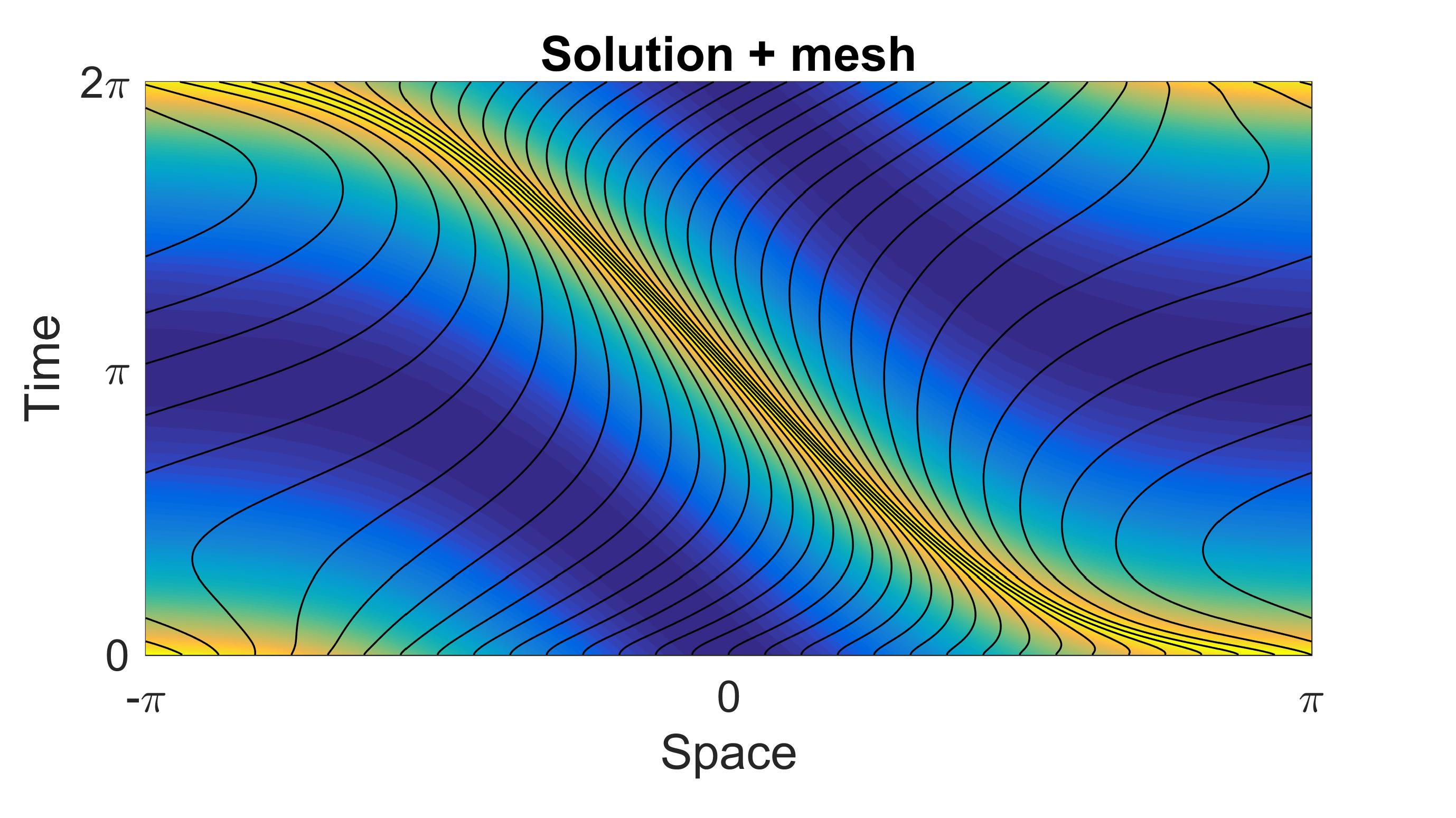}
	\caption{A solution to the heterogeneous advection equation.
		(Top)~Snapshots showing the formation of a sharp wave crest. The solution and its computed Hilbert transform are represented by solid and dashed lines respectively. Dots indicate mesh nodes.
		(Bottom)~A combined solution/mesh plot. Colours indicate the model solution (yellow high, blue low), and the trajectories of mesh nodes are shown as black lines. The mesh has been downsampled to $N=32$ nodes for clarity.
	}
	\label{fig:advection}
\end{figure}

The first model presented in this section is an advection equation with a heterogeneous sound speed:
\[
	u_t-u_xx_t = c(x)u_x,\qquad
	c(x) = \left[1+0.9\cos(x)\right]^{-1}.
\]
Recall that the usual time-derivative has been modified in the above expression to account for the movement of the mesh nodes.
This model describes linear wave propagation, with a propagation speed that is slower in the middle of the domain than the edges.
It is solved using a sinusoidal initial condition
\[
	u(0,x) = \cos(x-\pi),
\]
and the resulting wave is propagated until $t=2\pi$, when it has travelled the full length of the spatial domain and periodic wrapping begins to occur.
For this problem, the initial and final waveforms should be equal, and can be compared to measure the accuracy of a given simulation.
The heterogeneous sound speed causes the peak in the wave to sharpen as it propagates through the center of the domain, making an adaptive mesh beneficial.
A solution to this problem is depicted in Fig.~\ref{fig:advection}, computed using $N=64$ mesh nodes.
The snapshots in the upper subplot show the formation of the sharp wave crest, and the lower subplot shows the trajectories of the mesh nodes, which cluster densely around this crest.

\subsubsection{The viscous Burgers' equation}

The second model presented in this paper is the viscous Burgers' equation.
Within a moving mesh framework, it is given by
\begin{equation}
	\label{eq:burgers}
	u_t-u_xx_t = uu_x + \varepsilon u_{xx}.
\end{equation}
The nonlinear term causes this model to evolve shock fronts, with their severity controlled by diffusion at a rate given by $\varepsilon$.

\begin{figure}
	\includegraphics[width=\textwidth]{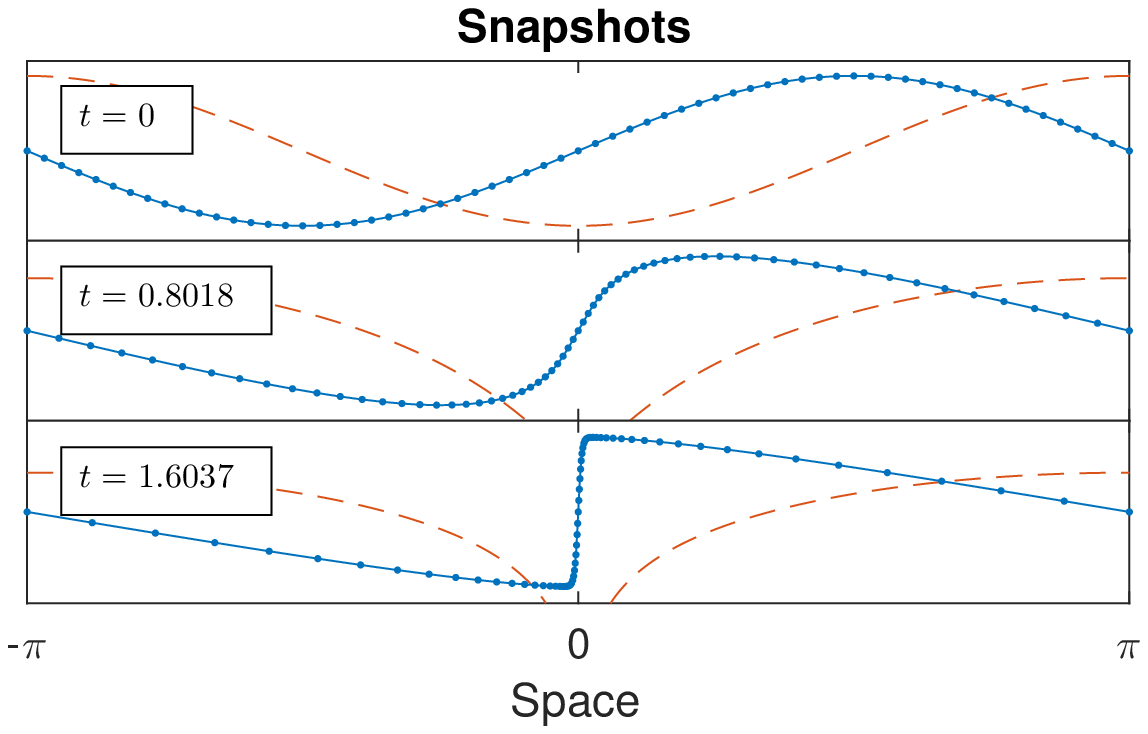}
	\includegraphics[width=\textwidth]{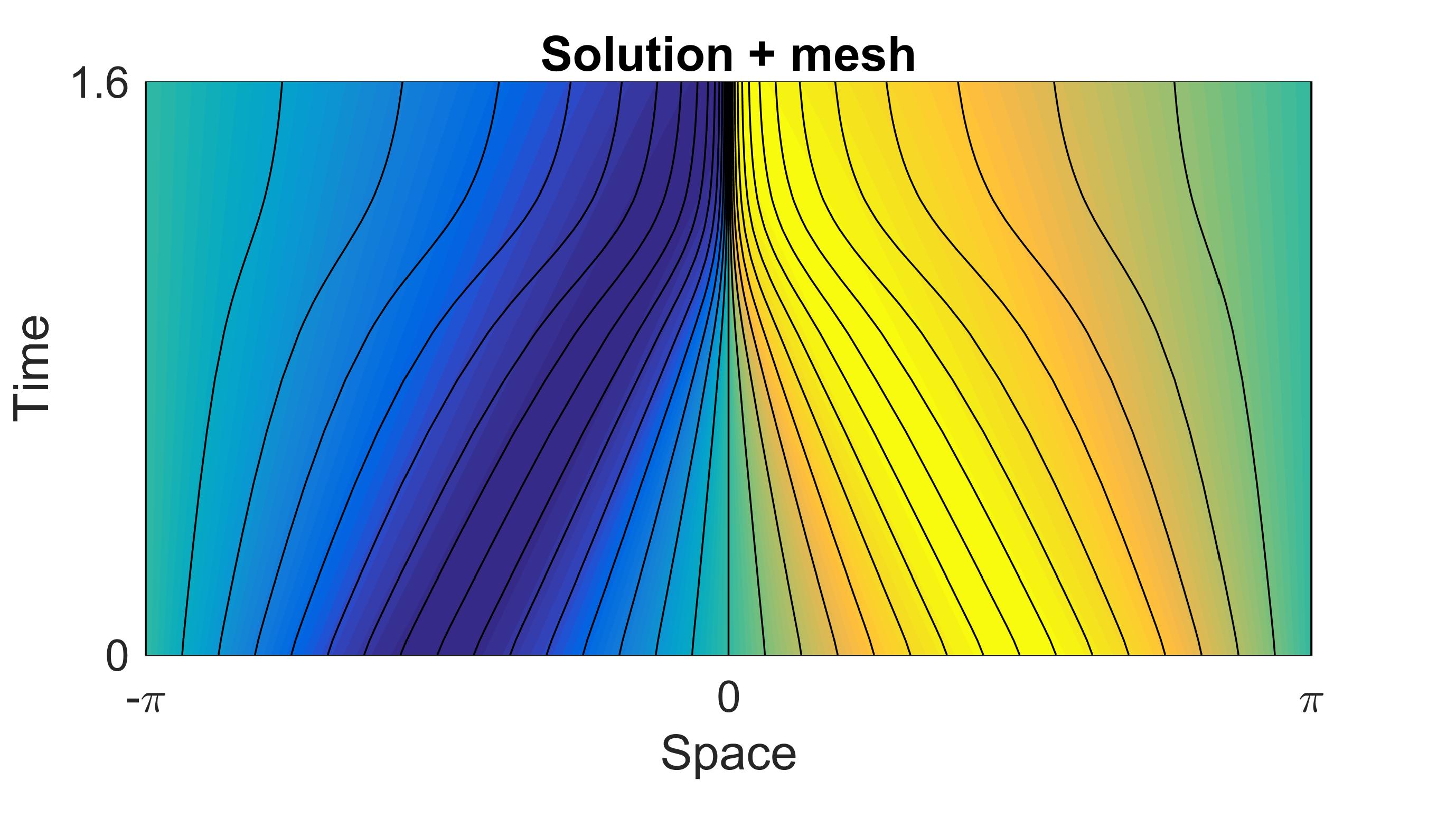}
	\caption{A solution to the viscous Burgers' equation.
		(Top)~Snapshots showing the formation of a steep, stationary shock front. The solution and its computed Hilbert transform are represented by solid and dashed lines respectively. Dots indicate mesh nodes.
		(Bottom)~A combined solution/mesh plot. Colours indicate the model solution (yellow high, blue low), and the trajectories of mesh nodes are shown as black lines. The mesh has been downsampled to $N=32$ nodes for clarity.
	}
	\label{fig:burgers}
\end{figure}

The first of two problems that use Burgers' equation exhibits a single, stationary shock front.
It is given by the following initial condition and diffusion coefficient:
\[
u(0,x) = \sin(x),\qquad
\varepsilon = 10^{-2}.
\]
(This problem is modified from that described in \cite{Guillard1988} to suit a domain of length $2\pi$.)
The simulation is terminated at $t=1.6037$, which is approximately when the shock front is steepest.
A solution to this problem is depicted in Fig.~\ref{fig:burgers}, computed using $N=64$ mesh nodes.
The snapshots in the upper subplot show the formation of the shock front, and the mesh node trajectories in the lower subplot smoothly converge around this shock front.

\begin{figure}
	\includegraphics[width=\textwidth]{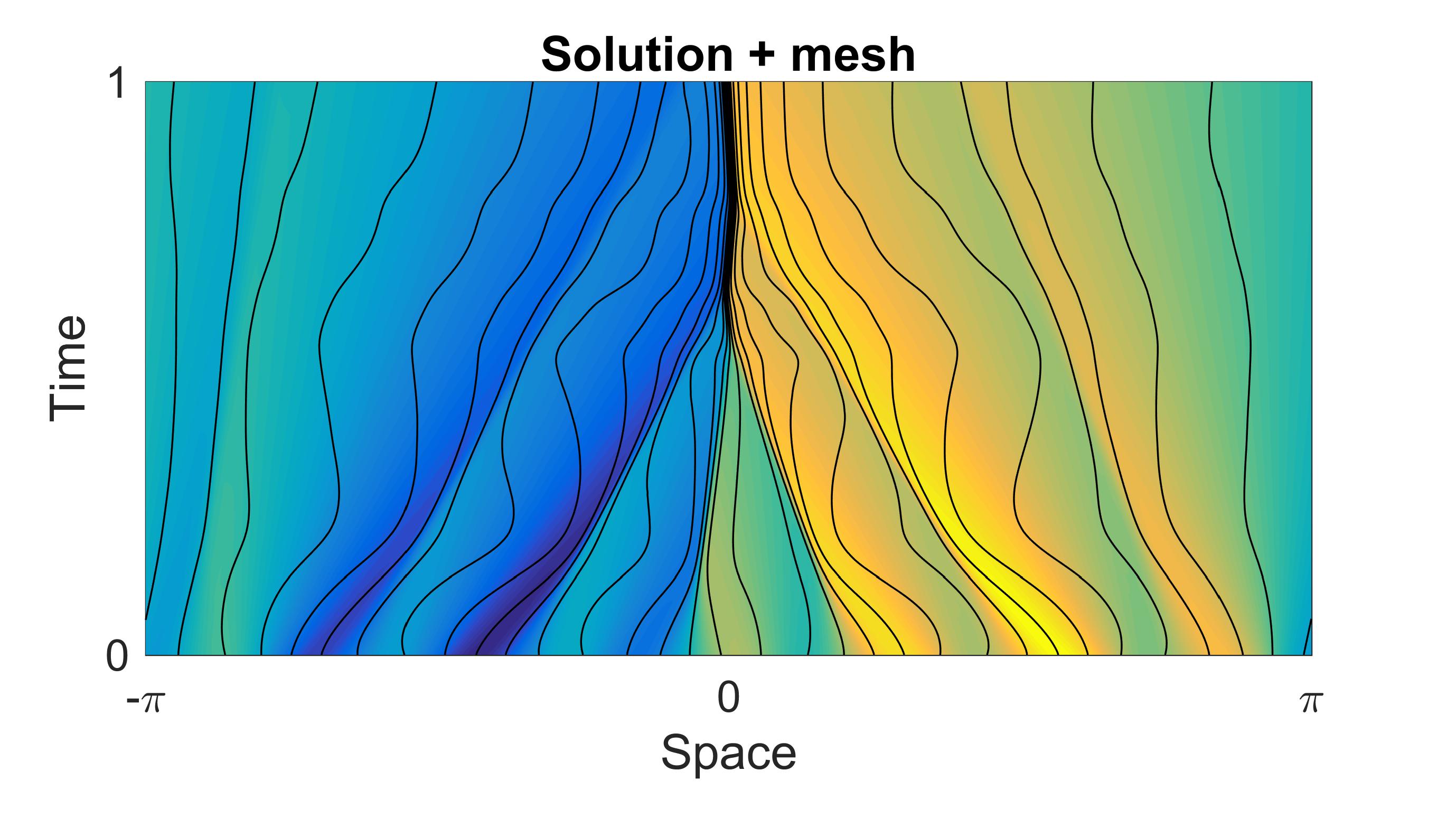}
	\caption{A combined solution/mesh plot for Burgers' equation. Colours indicate the model solution (yellow high, blue low), and the trajectories of mesh nodes are shown as black lines. The mesh has been downsampled to $N=32$ nodes for clarity. The mesh nodes track each shock front.
	}
	\label{fig:complicated-Burgers}
\end{figure}

The second problem that uses Burgers' equation exhibits seven propagating shock fronts that merge over time.
It is given by the following initial condition and diffusion coefficient:
\[
	u(0,x) = 2\sin(x)+\cos(7x),\qquad
	\varepsilon=10^{-2}.
\]
This simulation is terminated at $t=1$, when most of the shock fronts have merged, and the remainder have diffused significantly.
A solution to this problem is depicted in Fig.~\ref{fig:complicated-Burgers}, computed using $N=128$ mesh nodes.
The mesh nodes follow each wavefront smoothly, and become denser as the central shock front increases in severity.
This kind of problem was identified by Hale \cite{Hale2009b} as being problematic for singularity--based methods, which fail to generate a mesh when singularities coalesce (in this problem, each wavefront corresponds to a set of singularities).

\subsubsection{The Korteweg-de Vries equation}

\begin{figure}
	\includegraphics[width=\textwidth]{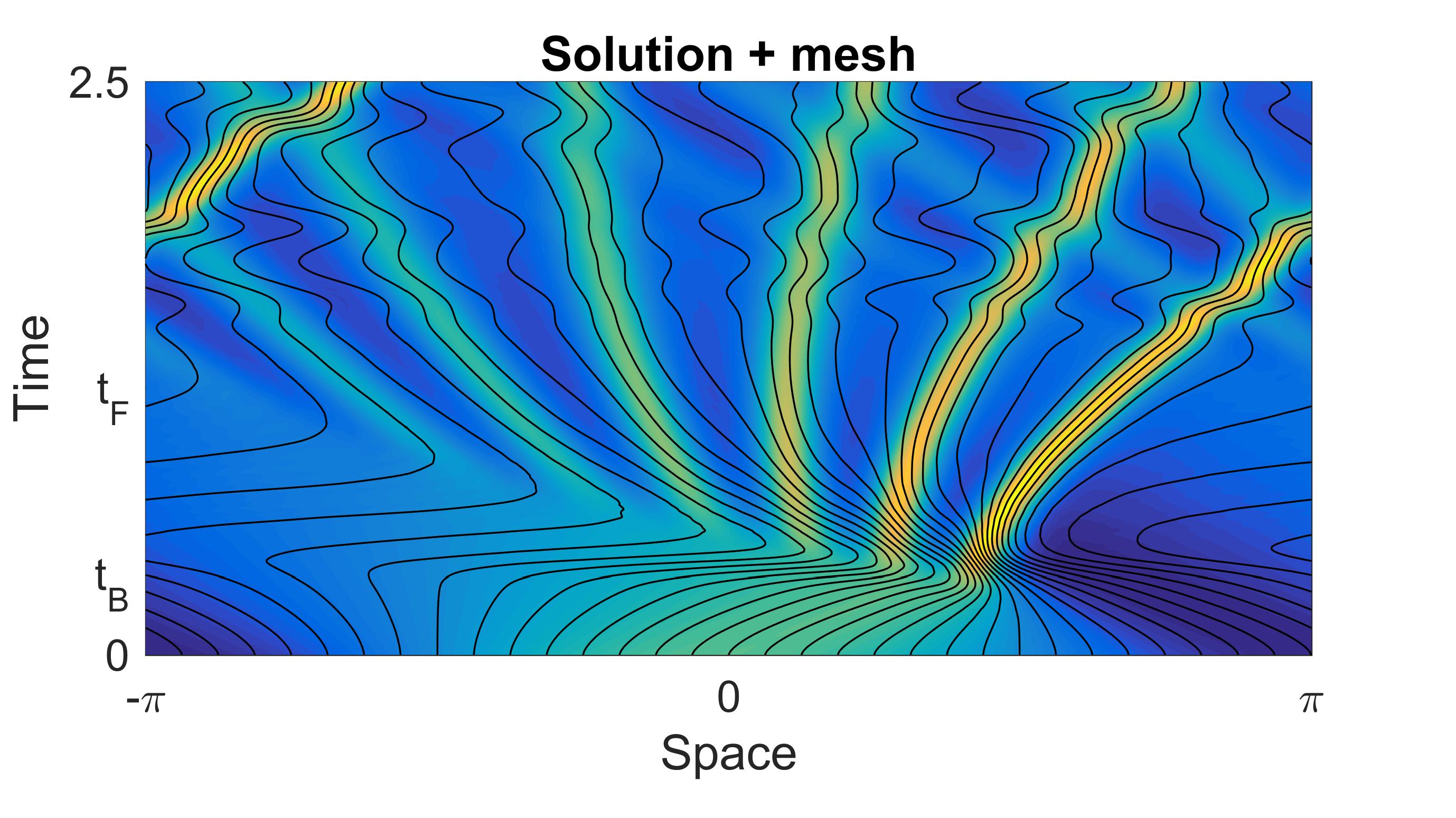}
	\caption{
		A combined solution/mesh plot for the Korteweg-de Vries equation. Colours indicate the model solution (yellow high, blue low), and the trajectories of mesh nodes are shown as black lines. The mesh has been downsampled to $N=32$ nodes for clarity. The mesh nodes compress around each soliton, and smoothly track them as they interact.
	}
	\label{fig:complicated-KdV}
\end{figure}

The third model presented in this paper is the Korteweg-de Vries equation, which combines nonlinear wave propagation with dispersion.
Within a moving mesh framework, it is given by
\begin{equation}
	\label{eq:korteweg}
	u_t-u_xx_t = \alpha uu_x + \beta u_{xxx}.
\end{equation}
The Korteweg-de Vries model admits solitons in its solutions.
These are waveforms whose size, shape, and velocity are constant provided they remain well separated.
Their speed is amplitude-dependent, and when two solitons interact the faster soliton is shifted forwards and the slower soliton is shifted back.
After interacting, solitons regain their original shape.
It is not clear that solitons are difficult to resolve since they appear visually smooth, but their analytic continuations are known to include singularities that limit convergence rates \cite{Tee2006b}.
The initial condition and parameters
\[
	u(0,x) = \sin(x),\qquad
	\alpha = -\pi,\qquad
	\beta = -(0.022)^2\pi^3,
\]
were chosen to match the problem presented in \cite{Zabusky1965}, modified to suit a domain of length $2\pi$.
The initial condition first steepens, before dispersion causes a number of solitons to form and propagate.
Soliton formation begins at approximately $t_B=1/\pi$, and completes at around $t_F=3.6t_B$.
After this, the solitons propagate until the periodic boundary conditions cause them to interact.
The simulation was terminated at $t=8t_B$.
A solution to this problem is depicted in Fig.~\ref{fig:complicated-KdV}, computed using $N=128$ mesh nodes.
The mesh nodes compress around each soliton, and continue to track them as they interact.

\subsection{Spectral convergence rates}

This section provides a performance evaluation for the bandwidth mesh density functions when applied to the periodic spectral method described in \S3.1.
They are judged by the rates at which approximated solutions to the problems in \S3.2 converge as the number of mesh nodes increases.
This illustrates the ability of meshes resulting from the bandwidth mesh density functions to reduce the tradeoff between computational resource usage and accuracy.

\begin{figure}
	\includegraphics[width=\textwidth]{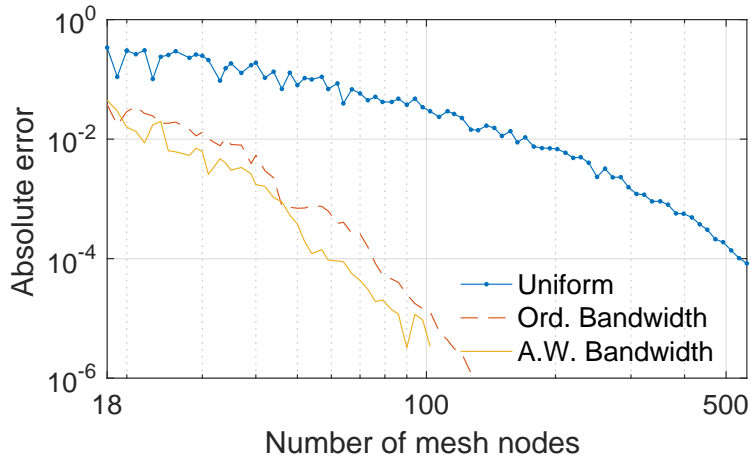}
	\caption{The error in the final waveform for the advection simulation depicted in Fig.~\ref{fig:advection} computed using a spectral moving mesh method. The solutions computed using the bandwidth mesh density functions exhibit much faster convergence rates that those that used a uniform mesh.}
	\label{fig:advection_performance}
\end{figure}

First, the advection problem was solved with a varying number of mesh nodes.
Three different mesh specifications were used: one uniform specification, and two based on the ordinary and amplitude-weighted bandwidth mesh density functions.
To evaluate their performance, the initial and final waveforms were interpolated onto 10,001 uniformly distributed mesh nodes and compared, since they should be equal to one another.
Figure~\ref{fig:advection_performance} depicts these results.
It is clear that both adaptive meshes produced error convergence rates which more than five times faster than those produced by the uniform mesh.
It also seems that there is a slight advantage to using the amplitude-weighted bandwidth mesh density function over the ordinary one.
An attempt was made to gather similar results using the arclength mesh density function
\begin{equation}
	\label{eq:arclength}
	\rho = \sqrt{|u_x|^2+1},
\end{equation}
but it proved difficult to produce a stable simulation.
This seemed to be due to the fact that the arclength function drops sharply at the crest of the wave, and is poorly resolved by the mesh that results from its use.
This difficulty may highlight the problem specificity of derivative-based mesh density functions when applied to spectral methods, since the feature that is difficult to resolve in this case is a point of high curvature.

\begin{figure}
	\includegraphics[width=\textwidth]{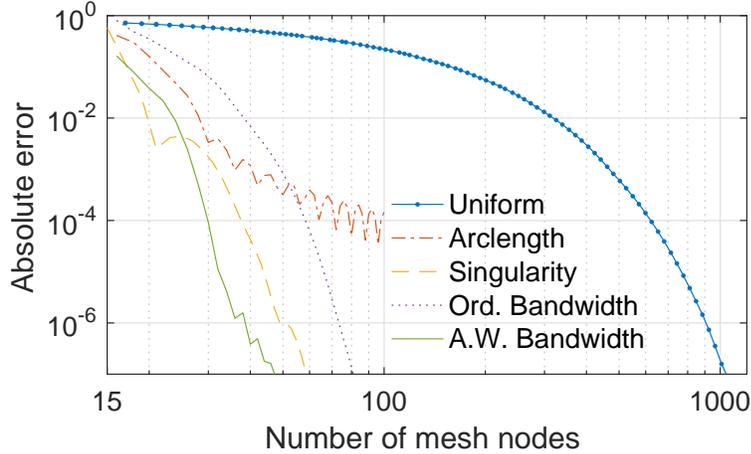}
	\caption{Convergence plots for solutions to the viscous Burgers' equation simulation depicted in Fig.~\ref{fig:burgers} using a spectral moving mesh method.
		The bandwidth mesh density functions clearly produce faster convergence rates than a uniform mesh specification or the arclength mesh density function, and converge at a similar rate to the singularity--based method of Tee et al.\ \cite{Tee2006a,Tee2006b,Hale2009a,Hale2009b}.
	}
	\label{fig:burgers_performance}
\end{figure}

A similar evaluation was performed for the first Burgers' equation problem (which exhibits one shock front).
A set of uniform mesh results were first generated for varying $N$, computed to near machine precision using Chebfun's \verb|spin| algorithm \cite{Driscoll2014}.
Results were then computed for four adaptive methods.
The first was the singularity--based method of Tee et al.\ \cite{Tee2006a,Tee2006b,Hale2009a,Hale2009b}.
This was applied with odd numbers of nodes in the range $N=15$ to $N=99$ (odd $N$ ensures a node at $x=0$ for their implementation).
The remaining results were computed using the spectral moving mesh method described in \S3.1, in conjunction with the arclength, ordinary bandwidth, and amplitude-weighted bandwidth mesh density functions.
These used even node numbers in the range $N=16$ to $N=100$.
All approximated solutions were interpolated onto 10,001 uniformly distributed mesh nodes and compared to the results computed using the \verb|spin| algorithm when the uniform mesh was at its densest.
Figure~\ref{fig:burgers_performance} depicts this comparison.
The uniform mesh produces very slow convergence for this problem.
The arclength mesh density function provides a significant improvement, but nonetheless converges far more slowly than the remaining methods.
The singularity--based method and the two bandwidth mesh density specifications clearly perform best.
These all produced error convergence rates that were more than an order of magnitude faster than those produced by a uniform mesh, and two to three times faster than those using the arclength mesh density function.
Once again, the amplitude-weighted bandwidth mesh density function outperforms its ordinary counterpart, this time by a larger margin.
It also outperforms the singularity--based approach, though by a smaller margin.

In addition to the results presented in Fig.~\ref{fig:burgers_performance}, results were gathered using the amplitude of the solution's analytic signal as a mesh density function.
It was found that this produced a small benefit over a uniform mesh.
This is likely because the feature of interest is a steep gradient, which produces an analytic signal with an infinite amplitude in the limit as the gradient's magnitude increases to infinity.
This may explain why the amplitude-weighted bandwidth mesh density function outperforms the ordinary bandwidth mesh density function for this problem.
In contrast, when the amplitude was used as a mesh density function for the advection problem, no benefit was found because the amplitude of the solution's analytic signal is approximately constant.
This illustrates the problem specificity of amplitude-weighting, and may motivate alternative mesh density functions for other applications, for instance one which includes an amplitude weighting for small amplitudes only.

\subsection{Finite-difference convergence rates}

This section provides a performance evaluation for the bandwidth mesh density functions when applied to the periodic finite-difference moving mesh method described in \S3.1.
This highlights the fact that the frequency considerations on which the bandwidth mesh density functions are based are relevant for piecewise polynomial interpolants as well as spectral ones.

\begin{figure}
	\includegraphics[width=\textwidth]{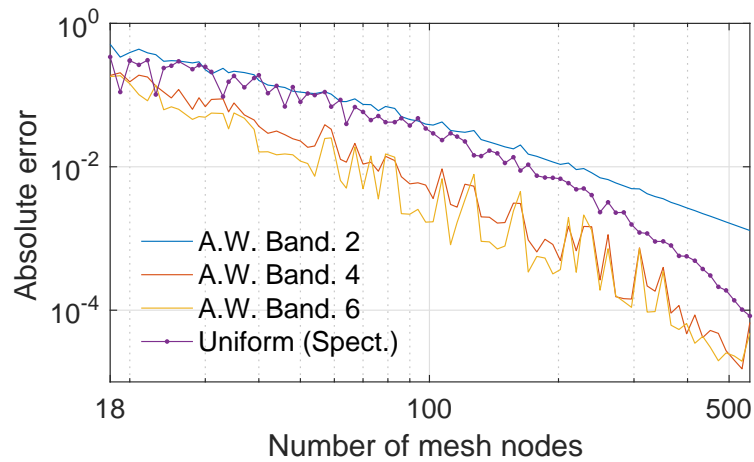}
	\caption{The error in the final waveform for the advection simulation depicted in Fig.~\ref{fig:advection} computed using
		centred, periodic finite-difference moving mesh methods.
		The numbers in the legend indicate the accuracy-order of the finite-difference method that was used.
		Adaptive meshes computed using the amplitude-weighted bandwidth mesh density function improve the performance of these finite difference methods to the point where they're comparable with a spectral method on a uniform mesh.
	}
	\label{fig:advection_performance_FD}
\end{figure}

First, the performance evaluation for the advection equation was repeated.
Three sets of results were computed using the amplitude-weighted bandwidth mesh density function, with each corresponding to a different accuracy-order for the finite differences.
Figure~\ref{fig:advection_performance_FD} depicts these results.
The spectral convergence rates have been replaced with algebraic ones, as expected, and as the accuracy-order of the finite-difference method increases, the accuracy improves.
Figure~\ref{fig:advection_performance_FD} also depicts the previous spectral results for a uniform mesh.
It is clear that the introduction of mesh adaptation can improve the performance of finite-difference methods to such an extent that they exceed that of the uniform spectral method.

\begin{figure}
	\includegraphics[width=\textwidth]{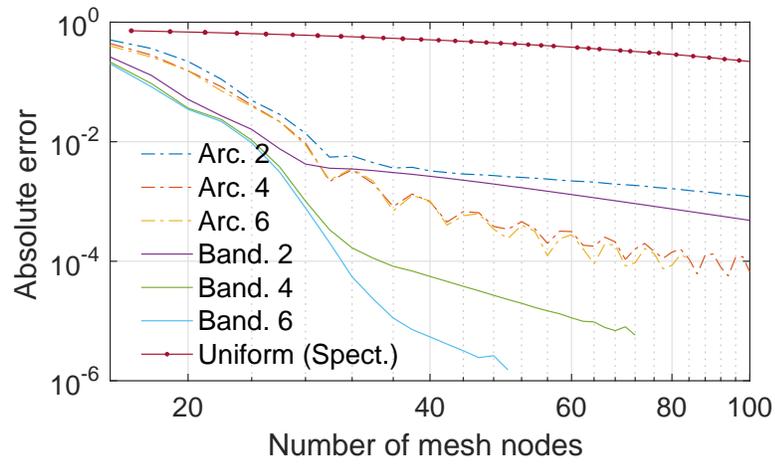}
	\caption{Convergence plots for solutions to the viscous Burgers' equation simulation depicted in Fig.~\ref{fig:burgers} using
		centred, periodic finite-difference moving mesh methods.
		The numbers in the legend indicate the accuracy-order of the finite-difference method that was used.
		The amplitude-weighted bandwidth mesh density function clearly produces faster convergence rates than the arclength mesh density function.
	}
	\label{fig:burgers_performance_FD}
\end{figure}

Second, the performance evaluation for Burgers' equation was repeated using the finite-difference moving mesh method.
In contrast to the results in \S3.3, only the arclength and amplitude-weighted bandwidth mesh density functions were examined.
Figure~\ref{fig:burgers_performance_FD} depicts these results.
The spectral convergence rates have been replaced with algebraic ones, as expected, and as the accuracy-order of the finite-difference method increases, the accuracy improves noticeably in almost all cases.
Comparing each finite-difference method, it is clear that the amplitude-weighted bandwidth mesh density function improve upon the results obtained using the arclength mesh density function significantly.
Figure~\ref{fig:burgers_performance_FD} also depicts the previous spectral results for a uniform mesh, and it is clear that the adaptive meshes drastically outperform a uniform mesh in this case.

\section{Summary and Discussion}

Two new mesh density functions have been presented in this paper, and applied to a number of model problems in acoustics.
They are called the bandwidth mesh density functions, and are derived from the local spatial frequency content of the model solution.
Their performance compares favourably with other approaches, and they are applicable to a wide range of problem types.
In particular, they have been shown to outperform the arclength mesh density function and to match the performance of singularity--based methods.
These numerical experiments were conducted using both a periodic spectral method and a periodic finite difference method, and good performance was observed for both.

The bandwidth mesh density functions presented in this paper are not the only ones that could be derived from consideration of the local spatial frequency content of a solution.
Possible alternative choices that could be made include:
\begin{itemize}
	\item Using a different order of spectral moment, or using a different spectral statistic altogether,
	\item Demodulating signals using an approach that doesn't involve the analytic signal \cite{Loughlin1996},
	\item Computing the local statistics using an explicit position--wavenumber power density,
	\item Adding an amplitude-weighting to the ordinary bandwidth mesh density function for small amplitudes only.
\end{itemize}
It is also pointed out in \cite{Subich2015} that not all frequency components in a solution will result in aliasing error in spectral collocation methods.
Hence, it may be useful to apply a high-pass filter as a preconditioning step before computing bandwidth mesh density functions.

The bandwidth mesh density functions can be extended to multidimensional problems using the Riesz transforms.
For a $d$-dimensional signal, the Riesz transforms yields $d$ components which can be combined with the original signal to form what is called a monogenic signal.
A similar approach to that presented in this paper may then be used to compute a multi-dimensional equivalent of the bandwidth mesh density functions.
The focus of future work will be on investigating this, and on applying the resultant monitor functions to large-scale medical acoustics simulations.

\section*{Acknowledgement}
The authors would like to thank the EPSRC for funding this work.

\section*{Appendices}
\appendix

\section{An algorithm for computing the Hilbert transform of a function from nonuniform samples}

The bandwidth mesh density functions require a complex-valued analytic signal to be defined from the real-valued solution to a model PDE.
In one dimension, this is typically done using the Hilbert transform, which can be defined in the frequency domain by
\[
\Hilb u = \Four^{-1}\left\{(-i\sign(k))\Four u \right\}.
\]
For equispaced samples, this is easily approximated using fast Fourier transforms.
For nonuniform samples, computing a Fourier transform is not so straightforward.
An alternative is to use its definition as a convolution
\begin{equation}
\label{eq:hilbert}
\Hilb u = \frac{1}{2\pi}\text{p.v.}\piint u(\tau)h(\tau-x)d\tau,\qquad
h(x) = \cot\left(\frac{x}{2}\right),
\end{equation}
where p.v. indicates the Cauchy principal value, and $h$ is the circular Hilbert kernel.

When computing this integrand, special consideration needs to be made for the singularity at $\tau=x$.
As described in \cite{Chen2009}, this singularity can be made removable by rewriting \eqref{eq:hilbert} as
\begin{equation}
\label{eq:hilbert2}
\Hilb u = \frac{1}{2\pi}\piint (u(\tau)-u(x))h(\tau-x)d\tau + \frac{u(x)}{2\pi}\text{p.v.}\piint h(\tau-x)d\tau.
\end{equation}
The singularity is now present only in the second integral; the first integral is instead of indeterminate form at $\tau=x$.
The singularity is easily dealt with by noting that the circular Hilbert kernel is antisymmetric and $2\pi$ periodic, meaning the Cauchy principal value of the second integral is equal to zero.

The indeterminate point in the first integral in \eqref{eq:hilbert2} is evaluated by computing the limit of the integrand as $\tau\to x$.
To do so, note that the Hilbert kernel can be obtained by making the Cauchy kernel, defined as $1/x$, $2\pi$-periodic:
\[
\cot\left(\frac{x}{2}\right) = 2\left[\frac{1}{x} + \sum_{n=1}^\infty\left(\frac{1}{x-2\pi n}+\frac{1}{x+2\pi n}\right)\right].
\]
Now let $t=\tau-x$ and $f(t)=u(\tau)-u(x)$.
From the first integrand in \eqref{eq:hilbert2}, write
\[
f(t)h(t) = 2f(t)\left[\frac{1}{t} + \sum_{n=1}^\infty\left(\frac{1}{t-2\pi n}+\frac{1}{t+2\pi n}\right)\right].
\]
As $t\to0$, each of the terms in the sum will cancel one another.
Thus,
\begin{equation}
\label{eq:limit}
\lim_{t\to0}f(t)h(t)=\lim_{t\to0}\frac{2f(t)}{t}.
\end{equation}
Because $f(0)=0$, this fraction is indeterminate.
It can be solved using L'Hospital's rule:
\begin{equation}
\label{eq:limit2}
\lim_{t\to0}f(t)h(t) = 2f'(0).
\end{equation}
Returning to the original variables, this is written as
\begin{equation}
\label{eq:final_limit}
\lim_{\tau\to x}\left[(u(\tau)-u(x))h(\tau-x)\right] = 2u'(\tau),
\end{equation}
which can be substituted into the left integrand in \eqref{eq:hilbert2} when it is solved.

\begin{subequations}
	\label{eq:hilbert_d}
	Finally, to compute the Hilbert transform, a change of variables is made from $x$ to $s$.
	Letting $\hat{u}=\Hilb u$, this gives
	\begin{equation}
	\hat{u}(x) = \frac{1}{2\pi}\piint F(x,\tau)\tau_sds,\qquad
	F = \left\{\begin{array}{l l}
	(u(\tau)-u(x))h(\tau-x) & \tau\neq x\\
	2u_\tau & \tau=x.
	\end{array}\right.
	\end{equation}
	This is solved using the trapezoid rule, discretised at a set of equispaced, discrete $s$ nodes.
	Let subscripts now denote the index of a variable, rather than differentiation.
	The variable $\tau$ is discretised equal to $x$, and so the resultant expression is
	\begin{equation}
	\hat{u}_i = \frac{1}{2\pi}\sum_j F_{i,j}\left(\pfrac{x}{s}\right)_j\Delta s,\qquad
	F_{i,j} = \left\{\begin{array}{l l}
	(u_j-u_i)h(x_j-x_i) & i\neq j\\
	2\left(\pfrac{u}{x}\right)_i & i=j.
	\end{array}\right.
	\end{equation}
\end{subequations}

\section{Analysis of the mesh density smoothing parameter}

\begin{figure}
	\includegraphics[width=\textwidth]{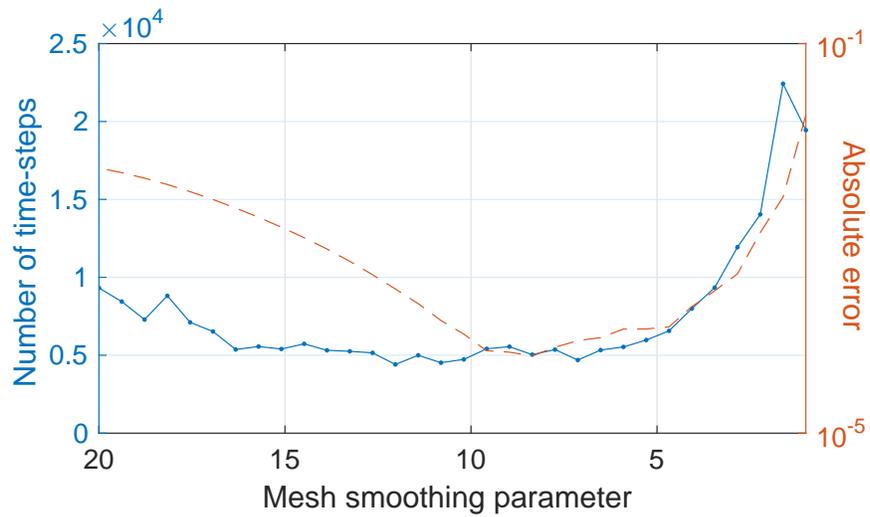}
	\caption{Error and number of timesteps for the advection simulation depicted in Fig.~\ref{fig:advection} with a varying smoothing parameter.
		Smoothing increases from left to right.
		Both the number of timesteps required (solid line) and the error (dashed line) decrease slightly as the mesh density function is increasingly smoothed, before turning around and increasing at a faster rate.
		Beyond the turning point, the mesh density function may become so smooth that the resulting mesh is no longer beneficial.
		Note that the minimum error is achieved for a smoothing parameter that is very close to the discretisation-dependent choice $\beta=(\Delta s\sqrt{2})^{-1}$ which equals 7.2 for this example.}
	\label{fig:advection_smooth}
\end{figure}

One free parameter in many moving mesh methods is the degree of smoothing that is applied to a computed monitor function prior to it's application within a MMPDE.
For the advection problem outlined in \S3.2.1 (and examined in \S3.3--3.4), the mesh smoothing parameter's effect was investigated by varying it from  $\beta=1$ to $\beta=20$.
Figure~\ref{fig:advection_smooth} depicts the error in the resulting solution, as well as the number of timesteps that were taken.
As the mesh density function is increasingly smoothed, both the number of timesteps required and the error slowly decrease, before turning and increasing at a faster rate.
This reflects the fact that the mesh density function has been smoothed to a point where it no longer produces a beneficial mesh.
The turning point in the error is close to the smoothing parameter choice mentioned in \S3.1, which is $\beta=(\Delta s\sqrt{2})^{-1}\approx7.2$ for this example.
Note that the scale of the changes in both the error and number of timesteps are quite small.
This indicates that the most important role of the smoothing parameter is to ensure that a solution can be computed stably, rather than optimising the speed or accuracy of the moving mesh method.

\bibliographystyle{ieeetrans} 
\bibliography{wise_bibliography}

\end{document}